\documentclass[a4paper,11pt]{iopart}

\usepackage[T1]{fontenc}
\usepackage[utf8]{inputenc}
\usepackage{iopams}
\usepackage{textcomp}

\usepackage{cite}

\usepackage{color}
\usepackage{graphicx}
\usepackage[hidelinks
]{hyperref}

\graphicspath{ {figures/} }

\usepackage{todonotes}

\usepackage{units}

\newcommand{\dd}{\mathrm{d}}
\newcommand{\micro}{\textrm{\textmu}}

\newcommand{\ee}{\mathrm{e}}
\newcommand{\ii}{\mathrm{i}}
\newcommand{\pp}{\mathrm{p}}
\newcommand{\sh}{\mathrm{sh}}

\newcommand{\nc}{n_{\mathrm{c}}}
\newcommand{\cs}{c_{\mathrm{s}}}
\newcommand{\lD}{\lambda_{\mathrm{D}}}
\newcommand{\ls}{l_{\mathrm{s}}}

\newcommand{\Cs}{{\mathrm{Cs}}}
\newcommand{\Cu}{{\mathrm{Cu}}}
\newcommand{\EE}{\mathcal{E}}
\newcommand{\MM}{\mathcal{M}}

\newcommand{\Smilei}{{\sf Smilei}}
\newcommand{\EPOCH}{\textsc{Epoch}}

\begin{document}

\noindent
\begin{minipage}{0.8\textwidth}
\vspace{-5em} \scriptsize {\it Originally published in Plasma
  Phys. Control. Fusion} {\bf 62} 085015, \quad\textcopyright~The
Authors, 2020
\\[1ex]
Original Content from this work may be used under the terms of the
\href{http://creativecommons.org/licenses/by/4.0/}{\textcolor{blue}{Creative
    Commons Attribution 4.0 licence}}. Any further distribution of
this work must maintain attribution to the author(s) and the title of
the work, journal citation and DOI.
\\[0.5ex]
doi:\href{https://doi.org/10.1088/1361-6587/ab9a62}{\textcolor{blue}{10.1088/1361-6587/ab9a62}}
\end{minipage}
\vspace{-2em}

\title[Collisional effects on shock dynamics in laser plasma]{%
  Collisional effects on the electrostatic shock dynamics in thin-foil
  targets driven by an ultraintense short pulse laser} %

\author{A~Sundstr\"{o}m$^{1}$, L~Gremillet$^{2,3}$, E~Siminos$^{4}$ and I~Pusztai$^{1}$}

\address{$^1$ Department of Physics, Chalmers University of
  Technology, SE-412\,96 G\"{o}teborg, Sweden} %
\address{$^2$ CEA, DAM, DIF, F-91297 Arpajon, France}
\address{$^3$ Universit\'e Paris-Saclay, CEA, LMCE, F-91680 Bruy\`eres-le-Ch\^atel, France}
\address{ $^4$ Department of Physics, Gothenburg University, %
  SE-412\,96 G\"{o}teborg, Sweden} %

\ead{andsunds@chalmers.se}

\begin{abstract}
We numerically investigate the impact of Coulomb collisions on the ion
dynamics in high-$Z$, solid density caesium hydride and copper
targets, irradiated by high-intensity
($I\approx\unit[2{-}5\times10^{20}]{Wcm^{-2}}$), ultrashort
(${\sim}\unit[10]{fs}$), circularly polarized laser pulses, using
particle-in-cell simulations.  Collisions significantly enhance
electron heating, thereby strongly increasing the speed of a shock
wave launched in the laser-plasma interaction.  In the caesium hydride
target, collisions between the two ion species heat the protons to
${\sim}\unit[100{-}1000]{eV}$ temperatures. However, in contrast to
previous work~(A.E.~Turrell \textit{et~al.},
2015~\emph{Nat.~Commun.}~{\bf 6}~8905), this process happens in the
upstream only, due to nearly total proton reflection.  This difference
is ascribed to distinct models used to treat collisions in dense/cold
plasmas.  In the case of a copper target, ion reflection can start
as a self-amplifying process, bootstrapping itself. Afterwards,
collisions between the reflected and upstream ions heat these two
populations significantly. When increasing the pulse duration to
$\unit[60]{fs}$, the shock front more clearly decouples from the laser
piston, and so can be studied without direct interference from the
laser.  The shock wave formed at early times exhibits properties
typical of both hydrodynamic and electrostatic shocks, including ion
reflection. At late times, the shock is seen to evolve into a
hydrodynamic blast wave.
\end{abstract}


\section{Introduction}
The use of lasers to accelerate ions is a field of intense
research~\cite{Macchi-etal_RevModPhys2013}, with many demonstrated or
envisioned applications, such as imaging of electromagnetic fields in
plasmas~\cite{Borghesi-etal_PoP2002,Romagnani-etal_PRL2005}, creation
of warm dense
matter~\cite{Patel-etal_PRL2003,Dyer-etal_PRL2008,Mancic-etal_PRL2010},
production of intense neutron sources~\cite{Roth-etal_PRL2013},
material
testing~\cite{Dromey-etal_NCommun2016,Barberio-etal_NCommun2018},
laboratory astrophysics~\cite{Higginson-etal_CommPhys2019}, and
ion-beam
therapy~\cite{Bulanov-etal_PhysLettA2002,Linz-Alonso_PRSTAB2007}.
Among the few laser-based ion acceleration mechanisms considered so
far, including the extensively studied, and particularly robust,
target normal sheath acceleration (TNSA), collisionless shock
acceleration (CSA) is of particular interest due to its potential to
produce a relatively narrowly peaked ion energy
spectrum~\cite{Denavit_PRL1992,Silva-etal_PRL2004,Romagnani-etal_PRL2008,Haberberger-etal_NPhys2012,Fiuza-etal_PRL2012,Pak-etal_PRAccelBeams2018}.
Collisionless shocks also play a role in particle energization in
astrophysical
plasmas~\cite{Karimabadi-etal_PoP2014,Dieckmann-etal_PPCF2017}.

As the shock front passes by, the plasma is rapidly compressed and
directional kinetic energy is converted into thermal energy. This can
take place either through collisional processes, such as in
hydrodynamic shocks~--~relevant in, e.g., inertial fusion
plasmas~\cite{Perkins-etal_PRL2009,Bellei-etal_PoP2014} and
relativistic laser-plasma experiments~\cite{Santos-etal_NJP2017}~--~or
collisionless mechanisms, involving longitudinal electrostatic fields
generated by space charge effects from shock
compression~\cite{Silva-etal_PRL2004}. %
Collisionless shocks can also hinge upon self-generated magnetic
fields, such as those resulting from the Weibel instability
\cite{Spitkovsky_ApJ2008,Lemoine-etal_PRL2019}, yet such shocks, of
turbulent character, develop at Mach numbers much larger than those of
the laminar electrostatic shocks that we shall address
here~\cite{Stockem-etal_SciRep2014}.
In relativistic laser-plasma interactions, electrostatic shocks can
arise either from the forward push exerted by the laser's
ponderomotive force (or ``laser piston'')~\cite{Silva-etal_PRL2004} in
the radiation pressure acceleration (RPA) regime, or from electron
pressure gradients in nonuniform plasmas~\cite{Fiuza-etal_PRL2012}.
While ``collisionless shocks'', as the name suggests, are sustained
through collective collisionless plasma processes, Coulomb collisions
may play a role in their dynamics. Indeed, a finite collisionality,
while affecting the shock, does not necessarily disrupt
it~\cite{CollShocks2019}.

Although the effect of collisions is often deemed negligible in
high-intensity laser-plasma interactions, due to the high particle
energies at play, it can become important when using solid or
near-solid density targets, especially if they contain elements of
high atomic numbers. %
In this paper, we consider two scenarios where collisions play an
important role: %
one has basic science interest while the other is relevant for high
energy density applications.  We also present cases with parameters in
between, to clarify how changes in laser and target parameters affect
the ion dynamics, and in particular the properties of the resulting
electrostatic shocks.  In all cases, we will consider a circularly
polarized femtosecond ($\unit[10{-}60]{fs}$) laser pulse.

The first case we consider is motivated by the work by Turrell,
Sherlock~\& Rose~\cite{Turrell-etal_NCommun2015} (hereafter referred
to as TSR), where it was reported that inter-species collisions in a
caesium hydride (CsH) target induce ultrafast collisional ion heating,
and essentially affect the shock dynamics. %
We find significantly different results compared to what is reported
by TSR, even though we study essentially the same physical setup. %
Importantly, we do not observe the occurrence of ultrafast proton
heating downstream of the shock, as most of the protons are reflected,
and as such, there is no appreciable inter-species friction in the
downstream.  As we will discuss, this discrepancy is likely due to a
different behaviour, at the high densities considered, of the
different collision algorithms employed by TSR and us.

The other case we address was first considered in a recent study of
ours~\cite{ElectronPaper2020} investigating ionization and collisional
electron heating effects in solid copper targets, relevant for
warm-dense-matter generation.  Here, we focus on the ion dynamics and
examine the impact on the generated shock of the increased electron
density in copper compared to CsH. We also assess the sensitivity of
the ion dynamics to the laser parameters and target thickness.

When using circular laser polarization, collisions dominate the
electron heating, which, in turn, results in the formation of a
stronger electrostatic shock compared to a purely collisionless
simulation. In the scenarios with copper, the evolution
of the shock is studied, from a hybrid hydrodynamical--electrostatic
shock, through a gradual dissipation of its energy, to the transition
to a hydrodynamical blast wave. In particular, the onset of shock
ion-reflection is found to be self-amplifying. Collisional friction
between the upstream and reflected ions heats the upstream ion
population, which enhances the fraction of reflected ions.

\section{Simulation study}
In this paper, we investigate two different target materials, caesium
hydride~(CsH) and pure copper~(Cu), both at their respective solid
densities. We perform one-dimensional~(1D) particle-in-cell~(PIC)
simulations with the \Smilei~PIC code~\cite{Smilei-paper}
(version~\texttt{4.1}), which has a collision module that has been
benchmarked~\cite{Perez-etal_PoP2012} in the
high-density/low-temperature regimes relevant for this paper.
In all cases, we use a circularly polarized~(CP),
$\lambda=\unit[800]{nm}$ wavelength laser with a Gaussian temporal
profile. The simulation box consists of 51200~cells over a length of
$\unit[20]{\micro{m}}$ (resolution $\Delta{x}=\unit[0.39]{nm}$), and a
$4^{\rm th}$ order interpolation shape function is employed.  The use
of a high-order shape function ensures good energy conservation
despite the Debye length in our collisionless simulation being
somewhat lower than the mesh size. The electrons are initialized at a
temperature of $T_{\ee,0}=\unit[1{-}10]{eV}$ and the ions at a
temperature of $\unit[0.1{-}1.0]{eV}$.

Both target materials contain a highly charged, $Z^{*}$, ion species,
such that the effect of collisions is significant. This high
collisionality turns out to be of crucial importance for the electron
heating. Since CP is used, the target electrons are energized through
inverse Bremsstrahlung rather than from the strongly inhibited
$j{\times}B$~\cite{Kruer-Estrabrook_PoF1985} or vacuum
heating~\cite{Bauer-Mulser_PoP2007,May-etal_PRE2011} mechanisms. In
our recent work~\cite{ElectronPaper2020}, we showed that collisional
electron heating produces well-thermalized electron populations with
temperatures in the ${\sim}\unit[1{-}10]{keV}$-range.

The use of the CsH target was inspired by the work by
TSR~\cite{Turrell-etal_NCommun2015}. As a target material, CsH could
be of interest for laser acceleration of protons since it contains
hydrogen volumetrically, like a plastic target. %
An advantage of this material over plastic, though, is the much higher
ionization degree ($Z^*$) that can be reached, hence enhancing
collisional effects. %
Although practically challenging, due to the high chemical reactivity
of CsH and difficulties in the target fabrication, it would, in
principle, be possible to use this material in an experiment.

The CsH target is composed of an equal number mixture of protons and
caesium ions. The charge state of the Cs ions is set to a fixed value
of $Z^*=27$, corresponding to full ionization of the three outermost
shells. The resulting quasi-neutral electron density is
${n_{\ee,0}=250\,\nc}$, where
$\nc=\epsilon_{0}m_{\ee}\omega^{2}/e^2\approx\unit[1.7\times10^{21}]{cm^{-3}}$
is the critical density ($\epsilon_0$ is the vacuum permittivity,
$m_{\ee}$ is the electron mass, $\omega$ is the laser frequency and
$e$ is the elementary charge), corresponding to a collisionless skin
depth of $\ls=\unit[8.0]{nm}$ which is well resolved. The target
thickness is $\unit[300]{nm}$, as in the simulations of TSR.

Copper, on the other hand, lacks the embedded protons but is, from a
practical standpoint, much more readily available as a target
material. Copper is also relatively highly charged, and hence presents
a collisionality comparable to CsH.  The lack of embedded
protons\footnote{%
  The copper is also modelled without any proton contamination layer
  on the surfaces. While such a contamination layer would affect the
  TNSA process, and somewhat the laser absorption, it is not expected
  to have a significant impact on the shock dynamics, that is the
  focus of this paper.} %
makes copper less suitable for volumetric proton acceleration, but its
high collisionality could be beneficial for other applications, such
as warm-dense-matter generation~\cite{ElectronPaper2020}.  In the
simulations, the copper ions are initialized with three fully ionized
atomic shells ($Z^*=27$), and at solid density (corresponding to
$n_{\ee,0}=1307\,\nc$ and $\ls=\unit[3.5]{nm}$). This choice is
informed by simulation results for a copper target including field and
collisional ionization processes, analyzed in
Ref.~\cite{ElectronPaper2020}, showing that the average $Z^{*}$
rapidly reaches this value, then it stagnates, due to a significant
jump in ionization energy beyond the three atomic shells. We found
that retaining the ionization dynamics has no significant impact on
the ion dynamics.

With the copper targets, two different target thicknesses and two
different laser parameters were considered. The thinner target is
$\unit[300]{nm}$ thick, as in the CsH simulations, which has the
advantage of quicker heating and homogenization compared to a thicker
target. The thicker ($\unit[2.5]{\micro{m}}$) target, on the other
hand, can be more suitable for warm-dense-matter applications: %
A high energy density will be maintained over a longer time since
hydrodynamic expansion takes longer to reach the interior of a thicker
target. We note that at the high densities and ionization degrees
considered here, the useful lifetime of the target can also be
affected by radiative losses, dominantly through Bremsstrahlung at the
temperatures of interest. We find, however, that for our parameters,
the radiative cooling time is typically of several picoseconds, so
that Bremsstrahlung losses should not greatly impact the plasma
dynamics during the integration time (${\le}\unit[1]{ps}$) of our
simulations. For the same reason, internal radiative energy transport
was also not modelled in the simulation.

We considered two different sets of laser parameters: an amplitude of
$a_0=15$ ($I\approx\unit[5\times10^{20}]{Wcm^{-2}}$) and
full-width-at-half-maximum (FWHM) duration of $\unit[10]{fs}$, as well
as $a_0=10$ ($I\approx\unit[2\times10^{20}]{Wcm^{-2}}$) and a FWHM
duration of $\unit[60]{fs}$. The former is used with both the CsH and
Cu thin targets, and the latter is used for both the thin and thick Cu
targets.  The use of thicker targets goes along with increased
integration times, allowing a larger number of fast particles to reach
the domain boundaries. To keep them inside the domain, the thicker
target is initialized with its front at $x=\unit[7.5]{\micro{m}}$
compared to the other targets located at $\unit[1]{\micro{m}}$.

For an accurate modelling of Coulomb collisions, employing the
relativistic PIC algorithm of~\cite{Perez-etal_PoP2012} (to be further
discussed in Sec.~\ref{sec:revisited}), a relatively high number of
particles per cell is needed. In the thinner target,
500~macro-particles per species per cell was used, while in the
thicker target, the particle number was reduced somewhat to
400~macro-particles per species per cell. Resolution tests, with
halved particle number or halved spatial resolution (with same total
number of particles), for the Cu thin target simulation show that the
simulations are numerically converged.

\section{Ion dynamics in the CsH target}
Motivated by the previous work by TSR, we performed a similar set of
simulations in CsH. However, despite virtually identical setups, our
results differ significantly from the ones by TSR.

\subsection{Comparison of collisional and collisionless results}

\begin{figure*}
\centering
\includegraphics{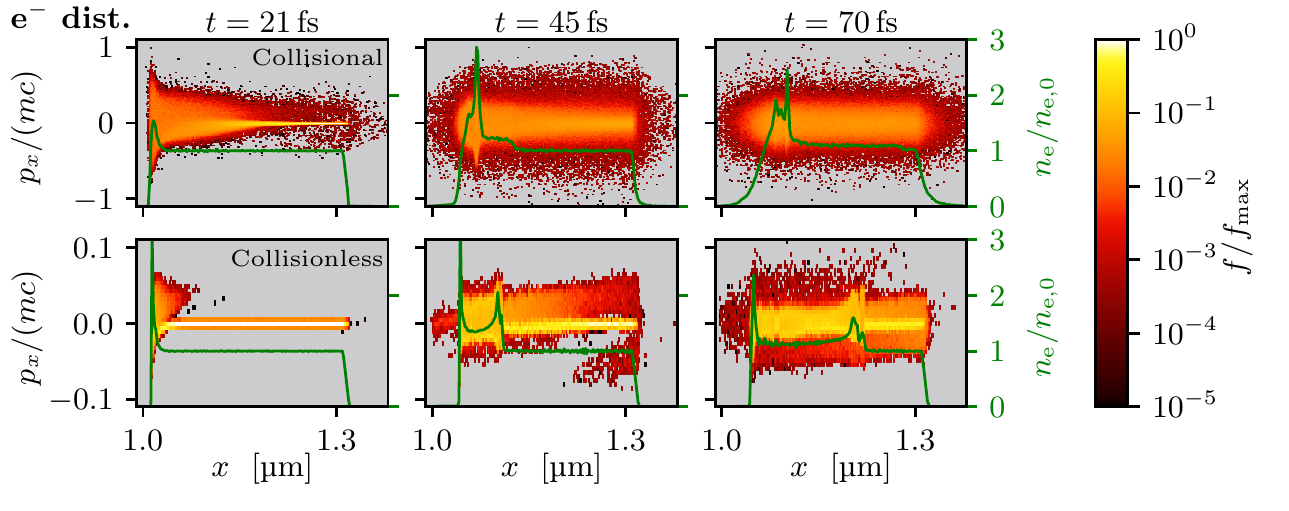}
\caption{ Electron distributions at~(left) and after the laser peak
  intensity~(middle and right), with (top) and without (bottom)
  collisions, using CP. Green curve: electron density. %
  Please note the different momentum scales for the collisional and
  collisionless simulations. %
}
\label{fig:CsH-thin_e-dist}
\end{figure*}

The primary effect of the strong target collisionality is to
significantly enhance electron heating through inverse
Bremsstrahlung~\cite{ElectronPaper2020}.  As an illustration of the
collisional electron heating, Fig.~\ref{fig:CsH-thin_e-dist} shows the
electron phase space of the collisional (top row) and collisionless
(bottom row) CsH simulations at three successive times: during peak
laser intensity at $t=\unit[21]{fs}$, right after the laser pulse has
ended at $t=\unit[45]{fs}$, and even later at $t=\unit[70]{fs}$. In
figures hereafter, the phase space distribution functions $f$ are
normalized to the maximum value of each respective \emph{initial}
Maxwellian distribution, $f_{\rm max}$.

The electrons in the target front layer are energized in the
transverse ($y$-$z$) plane by the laser electric field. Then
collisions scatter their momentum into the longitudinal direction, as
seen through the large spread in $p_{x}$ near the plasma front in the
$t=\unit[21]{fs}$ frame of the collisional distribution. Collisions
then entail a fast thermalization of the electrons to a Maxwellian
distribution, yielding a bulk temperature of
$T_{\ee}\approx\unit[10]{keV}$ that corresponds to an ion-acoustic
speed of
$\cs\approx(Z^{*}_{\Cs}T_{\ee}/m_{\Cs})^{1/2}\approx1.5\times10^{-3}c$, where $c$ is the speed of light in vacuum.

\begin{figure*}
\centering
\includegraphics{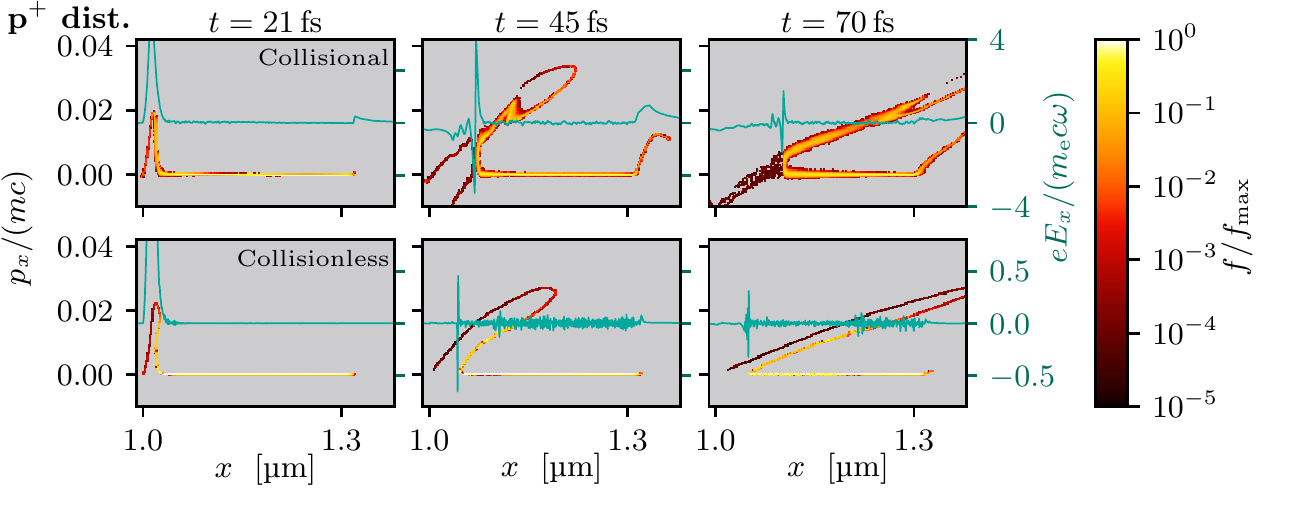}
\includegraphics{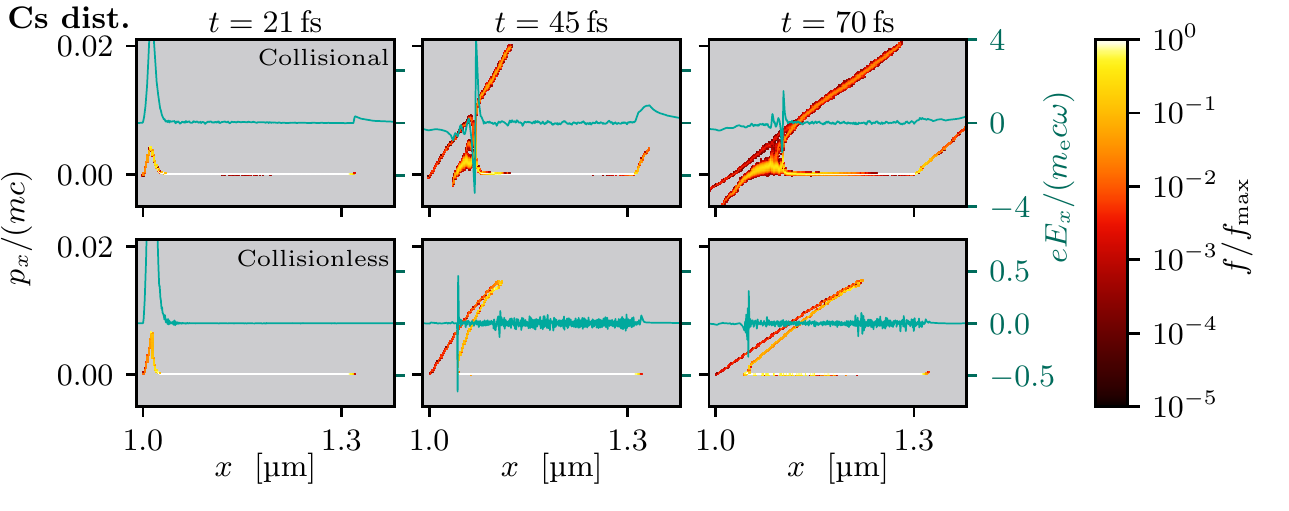}
\caption{Proton~(top frame) and caesium ion~(bottom frame)
  distributions in the $\unit[300]{nm}$ CsH target at peak laser
  intensity~($t=\unit[21]{fs}$) and after the pulse has
  passed~($t=\unit[45]{fs}$ and $\unit[70]{fs}$), with~(upper panels)
  and without~(lower panels) collisions, using CP. The longitudinal
  electric field is also plotted~(turquoise solid line, right axes). %
  Note the different electric field scales between the collisional and
  collisionless panels.  }
\label{fig:CsH-thin_i-dist}
\end{figure*}

The electron density is also indicated in
Fig.~\ref{fig:CsH-thin_e-dist} (green solid curve, right axis).
Compared to the collisionless case, the collisional simulation shows
smoother spatial structures likely due to a combination of higher
temperature, collisional dissipation and dispersion of non-linear
waves.  The Debye length is $\lD\approx\unit[1]{nm}$ and
$\lD\approx\unit[0.1]{nm}$ in the collisional and collisionless cases,
respectively. %
The collisional electron density profile also shows signs of an
electrostatic shock wave: a density jump moving away from the target
front is visible in the $t=\unit[45]{fs}$ and $\unit[70]{fs}$
panels. In the collisionless case, the density profile exhibits two
peaks in both time frames. The rightmost density jump is due to the
leading edge of the radiation-pressure-accelerated Cs ions
(Fig.~\ref{fig:CsH-thin_i-dist}), while the leftmost density peak
corresponds to an electrostatic shock, which, due to the low electron
temperature, is too slow for its propagation to be noticeable over the
displayed time and length scales.

In Fig.~\ref{fig:CsH-thin_i-dist}, the evolution of the ion
distributions in the collisional and collisionless CsH targets are
shown. The top frame shows the proton, the lower one the Cs ion phase
spaces, with the upper (lower) rows in both frames corresponding to
the collisional (collisionless) simulations, at times
$t=\unit[21]{fs}$, $t=\unit[45]{fs}$ and $t=\unit[70]{fs}$. At
$t=\unit[21]{fs}$, the difference between the collisional and
collisionless simulations is quite small; in both cases, the protons
and Cs ions are pushed by the laser piston. However, due to the lower
charge-to-mass ratio of the Cs ions compared to the protons, the Cs
ions react more slowly to the radiation pressure (RP) induced
electrostatic field (at $x\approx\unit[1]{\micro{m}}$) than the
protons, as seen by the almost four times higher velocity reached by
the protons ($p_x/mc\approx0.02$) at $t=\unit[21]{fs}$. Owing to the
short pulse duration ($10\, \rm fs$), the Cs ions do not have enough
time to react to RP before the pulse ends.

Also shown in Fig.~\ref{fig:CsH-thin_i-dist} is the longitudinal
electric field, $E_x$ (turquoise curve), normalized to
$m_{\ee}c\omega/e\approx\unit[4.013\times10^{12}]{V/m}$. %
The charge separation during the RPA phase creates a strong
longitudinal electric field, visible as a positive spike in $E_x$
close to $x=\unit[1]{\micro{m}}$ in the $t=\unit[21]{fs}$ panels.
Note that the peaks of the RP field are cut off in the display. The
collisionless RP field reaches a normalized amplitude of
$e E_x/(m_{\ee} c \omega)=8.6$, while the field in the collisional
simulation reaches only $5.6$. However, the RP field in the
collisional simulation has a wider spatial extent. When the electric
field is integrated, the potential drop across the RP field is
$e\phi\approx\unit[220]{keV}$ and $\unit[280]{keV}$ in the
collisionless and collisional cases, respectively. Thus, collisions do
not affect the RPA process significantly, as apparent from the
comparison of the collisional and collisionless panels at
$t=\unit[21]{fs}$ in Fig.~\ref{fig:CsH-thin_i-dist}.

With collisions, the electrostatic structure caused by RPA transforms
into an electrostatic shock, as evidenced by the single strong
oscillation of $E_x$ and modulations in the downstream ion
distribution in the $t=\unit[45]{fs}$ and $\unit[70]{fs}$ frames of
Fig.~\ref{fig:CsH-thin_i-dist}.  A close inspection of the
collisionless simulation reveals the same behaviour (although barely
visible in Fig.~\ref{fig:CsH-thin_i-dist}), indicating that an
electrostatic shock has also formed there.
However, due to the high electron temperature from collisional
heating, the shock is much stronger and faster in the collisional
case. In absolute units, the average shock velocity between
$t=\unit[45]{fs}$ and $\unit[70]{fs}$ was
$v_{\sh}/c\approx 4.3\times10^{-3}$ and
$v_{\sh}/c\approx 0.9\times10^{-3}$ in the collisional and
collisionless simulations, respectively.  Yet, the higher electron
temperature in the collisional target ($T_{\ee}\approx\unit[10]{keV}$
vs. $T_{\ee}\approx\unit[0.2]{keV}$) leads to a lower Mach number
($\MM\approx2.9$ vs. $\MM\approx4$).
The low shock speed in the collisionless simulation implies that the
shock-reflected ions have a significantly lower energy compared to
those originating from the initial burst of the RPA. In both the
collisional and collisionless cases, given its limited energy
reservoir provided by the ultrashort ($\unit[10]{fs}$) laser pulse,
the shock wave steadily loses its energy, as seen by the declining
field amplitude and the sloped reflected ion structure in the proton
and Cs phase spaces (i.e.\ the shocks are losing speed).

Another consequence of the efficient inverse Bremsstrahlung electron
heating is that the collisional simulation displays TNSA at the target
rear boundary, whereas it is virtually nonexistent in the
collisionless simulation, as evident at $t=\unit[70]{fs}$ in
Fig.~\ref{fig:CsH-thin_i-dist}. Due to the use of CP, the electrons
are weakly energized in the collisionless case, hence quenching TNSA.
In the collisional simulation, the TNSA protons attain energies
slightly lower than the RPA protons at the final simulation time.

In the collisional case, we also see that the reflected and upstream
proton and Cs ion populations are being significantly heated, in
contrast to their collisionless counterparts. By fitting Maxwellians
to the proton distribution in the range
$x=\unit[1.15{-}1.18]{\micro{m}}$ (close to, but still beyond direct
influence from the shock front) at time $t=\unit[70]{fs}$, the
upstream proton population is found to have already been heated to
$T_{\pp}^{\rm(u)}=\unit[120]{eV}$, while the reflected protons are at
a temperature of $T_{\pp}^{\rm(r)}=\unit[750]{eV}$. We recall that the
initial ion temperature was $\unit[0.1]{eV}$. Simulations in which
various types of collisions (e.g., proton--Cs or ion--electron) have
been selectively switched off (not shown here), reveal that the
heating of the reflected ions proceeds from their friction with the
background Cs ions, while the upstream ions are mainly collisionally
heated by the fast electrons.

\begin{figure}
\centering
\includegraphics{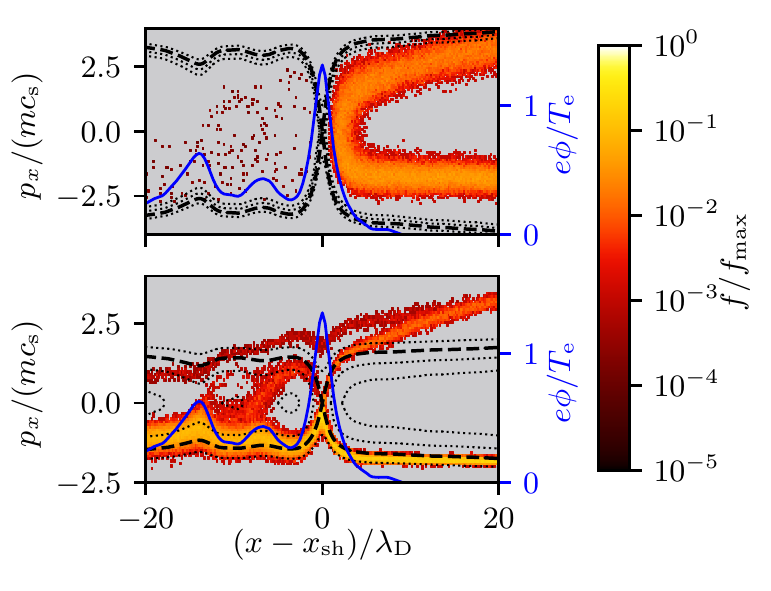}
\caption{Proton~(top panel) and Cs ion~(bottom panel) distributions in
  the shock frame of reference at $t=\unit[70]{fs}$, together with the
  shock electrostatic potential, $e\phi/T_{\ee}$ (blue solid line,
  right axes), using $T_{\ee}=\unit[10]{keV}$. Also shown are
  contours of constant energy, $\EE=mv^{2}/2+eZ(\phi-\phi_{\max})$
  (black, dashed or dotted lines). The black dashed line corresponds
  to $\EE=0$ at the potential peak, $\phi_{\max}$.}
\label{fig:CsH-thin_shock-dist}
\end{figure}

To get a more detailed picture of the vicinity of the electrostatic
shock front, close-ups of the proton (top) and Cs (bottom)
distributions, at $t=\unit[70]{fs}$, are displayed in
Fig.~\ref{fig:CsH-thin_shock-dist}. The distributions have been
shifted to the shock rest frame (at velocity
$v_{\sh}/c\approx3.1\times10^{-3}$), relative to the position of the
potential maximum, $x_{\sh}$; the velocities are normalized to the
ion-acoustic sound speed, $\cs$. The electrostatic potential,
$\phi(x)=-\int_{x_0}^{x}E_x(x')\,\dd{x'}$, where $x_0$ is such that
$\phi$ averages to zero in the range
$8\le(x-x_{\sh})/\lD\le10$\footnotemark{}, is also plotted (blue
line), along with corresponding constant energy contours (black dashed
or dotted lines). The black dashed line represents the constant energy
contour which has zero (shock-frame) kinetic energy at the peak of
$\phi$. This line is an approximate boundary between the reflected and
passing ions; in a steady state, this would be a separatrix. The top
frame shows that almost all protons are located within the reflected
region of phase space. Meanwhile, only around $5{-}10\%$ of the Cs
ions are reflected, and accordingly, the upstream Cs distribution
mostly lies below the passing--reflected boundary. The difference in
ion reflection between the two ion species is due to their different
charge-to-mass ratios~\cite{Pusztai-etal_PPCF2018}. %
\footnotetext{%
  In an idealized electrostatic shock, $\phi\to 0$ as $x\to\infty$.
  However, in practice, the electrostatic potential presents spatial
  variation even well upstream of the shock front, from sources other
  than the shock, which motivates this averaging procedure. The choice
  of the $x$-range to average over is somewhat arbitrary, but it is
  chosen reasonably close to the shock front, while sufficiently
  outside the shock width.}

The electrostatic potential is seen to oscillate downstream of the
shock (left side in Fig.~\ref{fig:CsH-thin_shock-dist}), which creates
regions of ion trapping. In a perfectly steady-state and collisionless
electrostatic shock, these regions would be empty, as there would be
no means for the ions to cross the separatrix. However, due to the
slowly decreasing amplitude and speed of the shock, the trapping
regions experience a steady influx of Cs ions. These adiabatic effects
are likely more important here than collisional
scattering~\cite{CollShocks2019}. While the Cs ions mainly enter the
trapping regions from the leftmost potential hump in
Fig.~\ref{fig:CsH-thin_shock-dist}, almost no protons pass the shock
front and hence only few protons ever enter the trapped region. The
protons trapped in those regions are mostly remnants of the protons
left behind the main RPA (seen to the left of the shock front in the
$t=\unit[45]{fs}$ frame of Fig.~\ref{fig:CsH-thin_i-dist}).

\subsection{Ultrafast ion heating revisited}
\label{sec:revisited}
The theoretical study of TSR~\cite{Turrell-etal_NCommun2015} predicts
that an ultrafast collisional ion heating may take place in plasmas
composed of light and heavy ion species.
This result is born out by 1D collisional PIC simulations performed
with the \EPOCH{}~\cite{Arber-etal_PPCF2015} code, considering a CsH
target almost identical to that in the current paper.  The authors
ascribe the observed ultrafast heating to collisional friction between
the protons and Cs ions as they experience a differential acceleration
in the electrostatic field of the shock.

The CsH setup presented in this paper is almost identical to that of
TSR~--~apart from a mere $1\%$ difference in the electron density, the
laser polarization and the increased resolution in our case. We have
also run a \Smilei{} simulation with exactly the same physical
parameters (including linear polarization and numerical resolution) as
TSR. As regards the ion dynamics, this simulation yields results
virtually identical to the CsH simulation presented
in~Fig~\ref{fig:CsH-thin_i-dist} (therefore, they are not presented
here separately). %
However, none of our simulations reproduce the main findings of TSR,
namely, the collisional downstream proton heating ascribed to
inter-species ion friction and the absence of ion reflection.  By
contrast, our simulations indicate that the collisional interaction
between the ion species does not inhibit the proton reflection; in
fact, as shown in e.g.\ Fig.~\ref{fig:CsH-thin_shock-dist}, nearly all
protons are reflected, and these are subsequently heated through
collisional friction through the ambient (upstream) Cs ions. The
proton heating is strongest in the \emph{reflected} ion population.

The PIC results of TSR are interpreted by a two-fluid model retaining
the momentum and energy moments of the Fokker-Planck equation,
assuming Maxwellian distributions. It provides steady-state
expressions for the longitudinal derivatives of the temperatures and
velocities of the two fluid species, which are then integrated over
the spatial width of the shock front.  The energy input to the system
comes from an electric field term representing the electrostatic shock
field. %
Importantly, the possibility of ion reflection is ruled out by
construction: protons are forced to pass through the barrier and gain
all the available potential energy, which is consistent with their
simulation results, but not with ours. %
In the collisionless case the protons should clearly be reflected due
to their higher charge to mass ratio than that of Cs. The only way to
avoid proton reflection is if a very strong friction between the two
species pulls the ions across the potential barrier. This, however,
requires a much stronger collisional coupling than what we observe.

Thus, we believe that the difference between TSR's results and ours is
(at least partly\footnote{%
  Modifications to the implementation of the
  Sentoku~\&~Kemp~\cite{Sentoku-Kemp_JCompPhys2008} collision model in
  \EPOCH{} over time make a direct comparison to TSR difficult.}) %
a consequence of the different collision algorithms used.  The version
of the \EPOCH~\cite{Arber-etal_PPCF2015} code used by TSR was equipped
with a collision module based on the algorithm proposed by
Sentoku~\&~Kemp~\cite{Sentoku-Kemp_JCompPhys2008} (SK), while
\Smilei{} employs the scheme developed by P\'{e}rez
et~al.~\cite{Perez-etal_PoP2012} (NYP), which generalizes the
Nanbu~\&~Yonemura
scheme~\cite{Nanbu_PRE1997,Nanbu-Yonemura_JCompPhys1998} to the
relativistic regime.  Both collision models are designed to reproduce
the Fokker--Planck limit, where small-angle collision events dominate,
which is relevant for high-temperature and/or low-density
plasmas. However, at the high plasma densities considered here, which
are susceptible to quantum degeneracy and coupled plasma effects,
corrections must be made to avoid unphysically high collision
frequencies\footnote{%
  It should be emphasized though, that these PIC simulations are
  intrinsically classical, and as such, a self-consistent treatment of
  quantum effects is clearly outside their scope. Thus the extensions
  of any binary collision model to dense/cold plasma regions are
  ad-hoc models designed to reproduce plasma-averaged collisional
  properties expected from advanced warm-dense-matter or
  condensed-matter models~\cite{Dharma-wardana-etal_PRE2017}.}. %
This is also a major point where the SK and NYP algorithms differ.

In the high-density/low-temperature regime, the SK model forces the
effective temperature of the interacting species to stay above the
Fermi temperature, in order to emulate the Fermi-degenerate
regime. This leads to the maximum collision frequency
$\hat{\nu}_{\alpha\beta}^{\rm(SK)}=m_{\ee}Z^{*}_{\beta}e^{4}\log\Lambda
/(12\pi^3\epsilon_{0}^2\hbar^3)$ %
between two particles of species $\alpha$ and
$\beta$~\cite[eq{.}~(10)]{Sentoku-Kemp_JCompPhys2008}.  By contrast,
drawing from the prescription of Ref.~\cite{Lee-More_PoF1984} for
coupled plasmas, the NYP model applies a lower bound on the
collisional mean free path, which can never get smaller than the mean
inter-particle distance $r_{\beta}\sim (4\pi n_{\beta}/3)^{-1/3}$. %
This yields the saturated collision frequency
$\hat{\nu}_{\alpha\beta}^{\rm(NYP)}=(4\pi n_{\beta}/3)^{1/3}
(T_{\alpha}/2m_{\alpha})^{1/2}$~\cite[sec.~I-C]{Perez-etal_PoP2012}.

At the considered ion density
($n_{\Cs}=\unit[1.5\times10^{28}]{m^{-3}}$) and a representative ion
temperature of $T_{\ii}=\unit[100]{eV}$, we find
$\nu_{\pp\,\Cs}^{\rm(Spitzer)}\approx\unit[1.5\times10^{15}]{s^{-1}}$,
and $\lambda_{\pp\,\Cs}\approx\unit[0.06]{nm}$ that is significantly
smaller than the inter-atomic distance ${\sim}\unit[0.2]{nm}$. Thus the
dense-plasma limit of NYP should hold under such conditions but not
the degenerate SK limit. One thus obtains that
$\nu_{\pp\,\Cs}^{\rm(SK)} = \nu_{\pp\,\Cs}^{\rm(Spitzer)}
\approx\unit[1.5\times10^{15}]{s^{-1}}$ is more than $5$ times larger
than the dense-plasma value
$\nu_{\pp\,\Cs}^{\rm(NYP)}\approx\unit[2.7\times10^{14}]{s^{-1}}$.
This discrepancy is only strengthened when considering Cs--Cs
collisions. Again at $n_{\Cs}=\unit[1.5\times10^{28}]{m^{-3}}$ and
$T_{\ii}=\unit[100]{eV}$, one finds
$\nu_{\Cs\,\Cs}^{\rm(Spitzer)}\approx\unit[6.6\times10^{16}]{s^{-1}}$,
which is over three orders of magnitude larger than the dense-plasma
NYP value,
$\nu_{\Cs\,\Cs}^{\rm(NYP)}\approx\unit[2.4\times10^{13}]{s^{-1}}$.
Regarding the electron--Cs ion collisions, one has
$\nu_{\ee\,\Cs}^{\rm(SK)}=\nu_{\ee\,\Cs}^{\rm(Spitzer)}\approx\unit[6.3\times10^{16}]{s^{-1}}$
at $T_{\ee}=\unit[100]{eV}$, but this corresponds to a mean free path
$\lambda_{\ee\,\Cs}\approx\unit[0.06]{nm}\ll r_{\Cs}$, so again the
dense-plasma limit applies, which gives
$\nu_{\ee\,\Cs}^{\rm(NYP)}\approx\unit[1.2\times10^{16}]{s^{-1}}$. The
difference is even larger at the lower temperatures associated with
the early-time interaction.

Moreover, the SK and NYP schemes handle colliding particles with
non-equal statistical weights differently, which impacts the accuracy
of energy conservation. However, that is likely not the cause of the
diverging simulation results, since the number of computational
particles is large in both cases, so as to limit statistical noise.

A recent simulation study~\cite{Bhadoria-etal_arxiv2019} of dense
($n_{\ee}=60\nc$) plasmas driven at relativistic laser intensities,
comparing the results of the SK and NYP\footnote{%
  Since version~\texttt{4.17} (June 2019) \EPOCH{} has the full NYP
  algorithm implemented as well.} %
modules in \EPOCH{} and the NYP module of \Smilei, confirms that the
SK model indeed results in stronger effective collisionality. In
addition, a good agreement between \EPOCH{} and \Smilei{} was found
when both employed the NYP algorithm.

Which of these two treatments of collisions in dense/cold plasmas is
more physically correct is still a debated issue. Therefore, along
with further numerical investigation, experimental verification should
be sought for in order to determine the parameter regions of validity,
and accuracy, of the two collision algorithms. Our results suggest
that such differentiation between the algorithms is possible using
laser-plasma experiments in multi-species, dense plasmas, such as the
CsH case presented here. A good benchmarking test would be to compare
ion energy spectra in cases where collisions are sufficiently strong
to suppress ion reflection according to SK but not according to
NYP. Such experiments might need to control the target density profile
on the rear side, e.g. through laser ablation, in order to suppress
TNSA, and make the shock accelerated ion population clearer. A
potentially suitable experiment has recently been
performed~\cite{Pak-etal_PRAccelBeams2018}, but it would require
further investigation to see whether the accuracy of the two models
can be assessed from the obtained data (which clearly showed ion
reflection).

\section{Ion dynamics in copper targets}

\begin{figure*}
\centering
\includegraphics{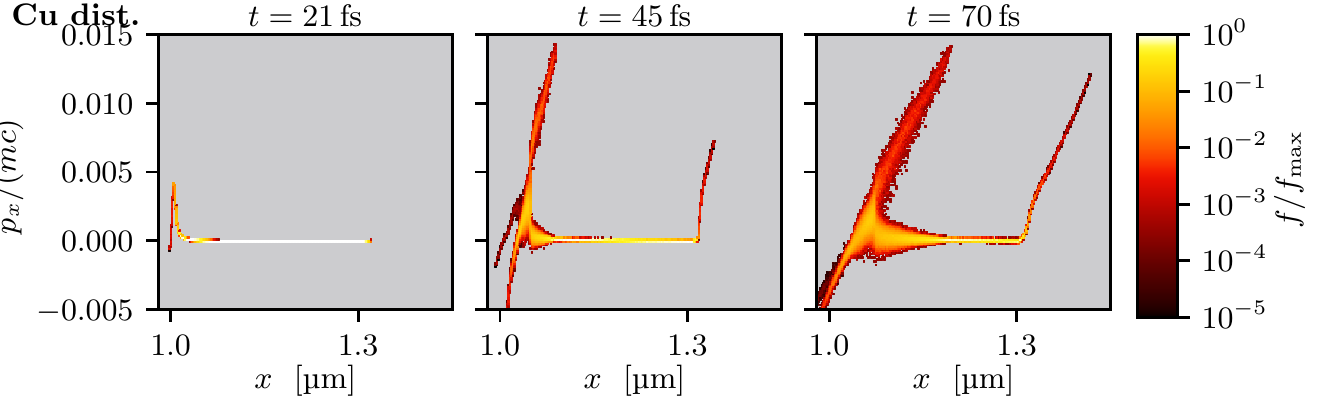}
\caption{Copper ion phase-space distribution in the $\unit[300]{nm}$
  thick Cu target from the collisional simulation, at times
  $t=\unit[21]{fs}$, $t=\unit[45]{fs}$ and $\unit[70]{fs}$.
}
\label{fig:Cu-thin_i-dist}
\end{figure*}

We will now turn to the pure copper simulations, first considering
similar target and laser parameters to the CsH case, and subsequently
changing these parameters one by one. The two main differences
compared to CsH are the lack of multi-species effects and the
${\sim}5$ times higher electron density (assuming
$Z^{*}=27$). However, just as in the CsH target, the primary effect of
collisions in the Cu plasma is the inverse Bremsstrahlung-type
electron heating. The bulk electrons are heated to
$T_{\ee}\approx\unit[3.7]{keV}$, corresponding to a sound speed of
$\cs\approx1.3\times10^{-3}c$.

Figure~\ref{fig:Cu-thin_i-dist} shows the collisional Cu ion
phase-space distribution, at peak laser intensity ($t=\unit[21]{fs}$),
close after the laser irradiation ($t=\unit[45]{fs}$), and even later
in time ($t=\unit[70]{fs}$). %
Similarly to Cs, the Cu ions have a rather low charge-to-mass ratio
($Z^{*}/A=0.42$) and do not have time to fully respond to the laser
piston during the short-pulse irradiation.  Again, the initial
perturbation from the laser piston transforms into an electrostatic
shock, yet it is losing energy faster than in the CsH target.

Since the copper plasma does not contain any protons, all of the
reflected charge, needed to sustain the shock, consists of Cu
ions. Owing to their high charge ($Z^{*}=27$), the collisional
interaction between the reflected and the upstream ions is stronger
than in the collisional CsH case, resulting in a noticeable heating of
these two populations, as seen in the collisional Cu ion distributions
at $\unit[45]{fs}$ and $\unit[70]{fs}$ in
Fig.~\ref{fig:Cu-thin_i-dist}. Some heating is observed in the
collisional proton and Cs ion distributions of
Fig.~\ref{fig:CsH-thin_i-dist} as well, but significantly weaker than
in the copper plasma.

\begin{figure*}
\centering
\includegraphics{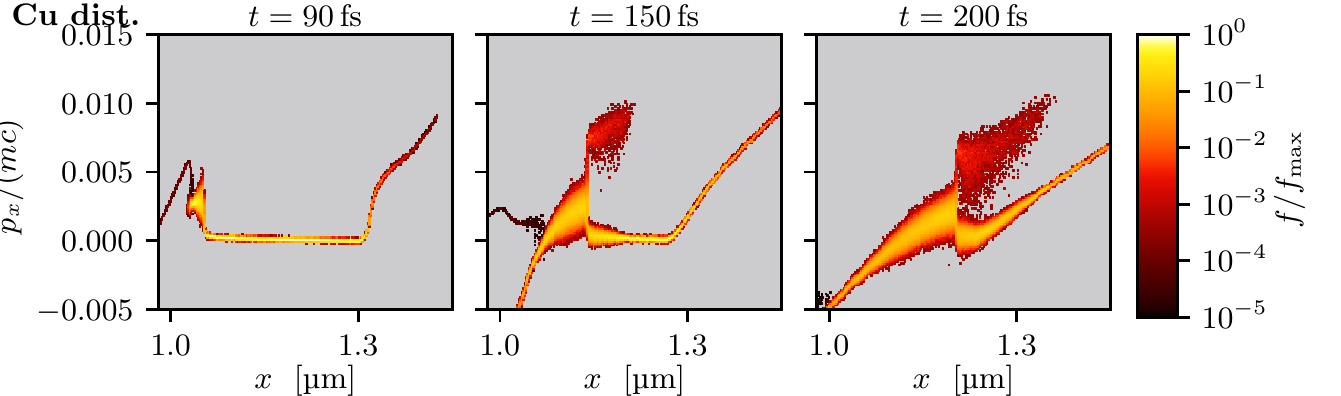}
\caption{Copper ion phase-space distribution in the $\unit[300]{nm}$
  thick Cu target from the collisional simulation, with an $a_0=10$
  and $\unit[60]{fs}$ duration laser pulse. The distribution is shown
  at times $t=\unit[90]{fs}$, $t=\unit[150]{fs}$ and $\unit[200]{fs}$.
}
\label{fig:Cu-thin_long-laser_i-dist}
\end{figure*}

\begin{figure}
\centering
\includegraphics{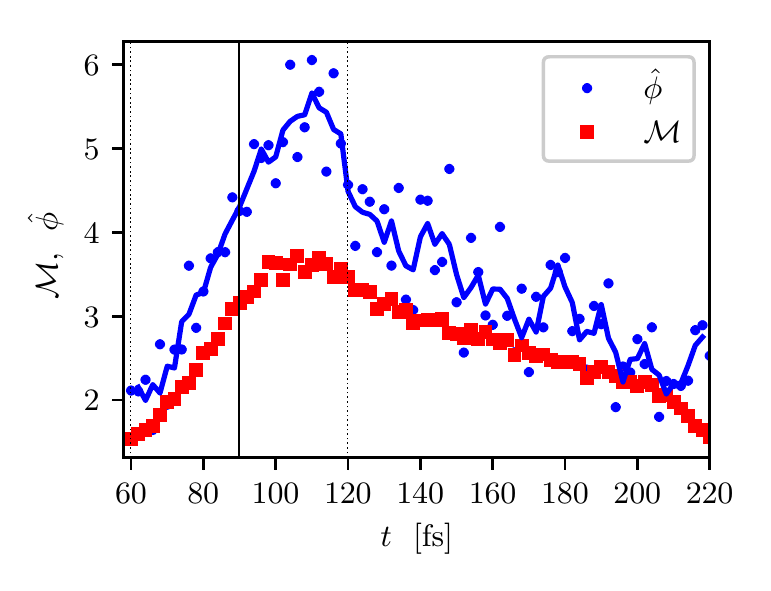}
\caption{Temporal evolution of the normalized potential drop across
  the shock front, $\hat{\phi}=e\phi/T_{\ee}$, and shock Mach number,
  $\MM$, for the thin copper target, long pulse collisional
  simulation. The blue line represents a moving average (over three
  data points) of $\hat{\phi}$. The vertical lines indicate the time of
  peak (solid) and half (dotted) laser intensity. For both the 
  normalization of the potential and for the ion acoustic sound speed, 
  an electron temperature of $T_{\ee}=\unit[4]{keV}$ was used as a 
  representative value. }
\label{fig:Cu-thin_long-laser_shock-M-dphi}
\end{figure}

We have also studied a scenario wherein the copper plasma is
illuminated by a laser pulse of longer duration ($\unit[60]{fs}$ FWHM)
and lower intensity ($a_0=10$). The Cu ion phase-space distribution
from the collisional simulation is displayed in
Fig.~\ref{fig:Cu-thin_long-laser_i-dist}, shown at times
$t=\unit[90]{fs}$ (at peak laser intensity), $t=\unit[150]{fs}$ (close
to the end of the pulse) and $t=\unit[200]{fs}$. Like in the two
previous setups, an electrostatic shock is generated. It forms out of a
perturbation that detaches from the laser piston already as early as
$t\approx\unit[60]{fs}$, before the pulse has reached half its maximum
intensity. It displays electrostatic shock-like properties, such as a
sharp rise in lab-frame ion velocity in conjunction with a steep
electrostatic potential barrier, but it lacks any ion reflection, as
seen in the $t=\unit[90]{fs}$ frame of the collisional simulation.

Figure~\ref{fig:Cu-thin_long-laser_shock-M-dphi} shows the evolution
of the normalized\footnote{%
  Using a fixed value of $T_{\ee}=\unit[4]{keV}$, derived from
  Maxwellian fits to the electron energy spectrum (whole plasma). The
  measured electron temperature stays fairly close to this value
  during the entire duration of the pre-shock and the electrostatic
  shock. } %
potential jump $\hat{\phi}=e\phi/T_{\ee}$ and Mach number $\MM$ of the
shock from the time of detachment from the laser piston to its
demise. The transition to a fully developed, ion-reflecting,
electrostatic shock occurs when $\hat{\phi}\gtrsim\MM^2/2$, which is
at around $t\approx\unit[90]{fs}$. The longer pulse duration and more
gradual increase in intensity, detaches the onset of shock reflection
from RPA.

The peaks in $\hat{\phi}$ and $\MM$ are followed by a more gradual
decrease in the Mach number and in the shock potential peak, starting
at around $t\approx\unit[110]{fs}$. The vertical lines in
Fig.~\ref{fig:Cu-thin_long-laser_shock-M-dphi} represent the time of
peak (solid) and half (dotted) laser intensity. The peaks thus occur
before the laser intensity has halved. The delayed peaks in shock
speed and potential relative to the peak laser intensity likely
originate from the fact that the interaction has reached a stage where
the laser is no longer able to supply more power than the energy
dissipation rate of the shock.

The reflection of ions appears as a process bootstrapping
itself. After the first few ions have been reflected, collisional
heating between the upstream and reflected ions cause a broadening of
the longitudinal momentum distribution of the upstream ions, leading
to more ions entering the reflected region of phase-space. This
upstream heating is seen in the collisional $t=\unit[150]{fs}$ frame
of Fig.~\ref{fig:Cu-thin_long-laser_i-dist}. %
Towards the end of the simulation, the upstream and reflected ion
populations start to merge into each other, after which the
determination of the shock speed relative to the upstream population
becomes unreliable. The shock ends somewhat abruptly when it
collides with the rarefaction wave emanating from the back of the
target, which occurs at roughly $t\approx\unit[250]{fs}$.

\begin{figure*}
\centering
\includegraphics{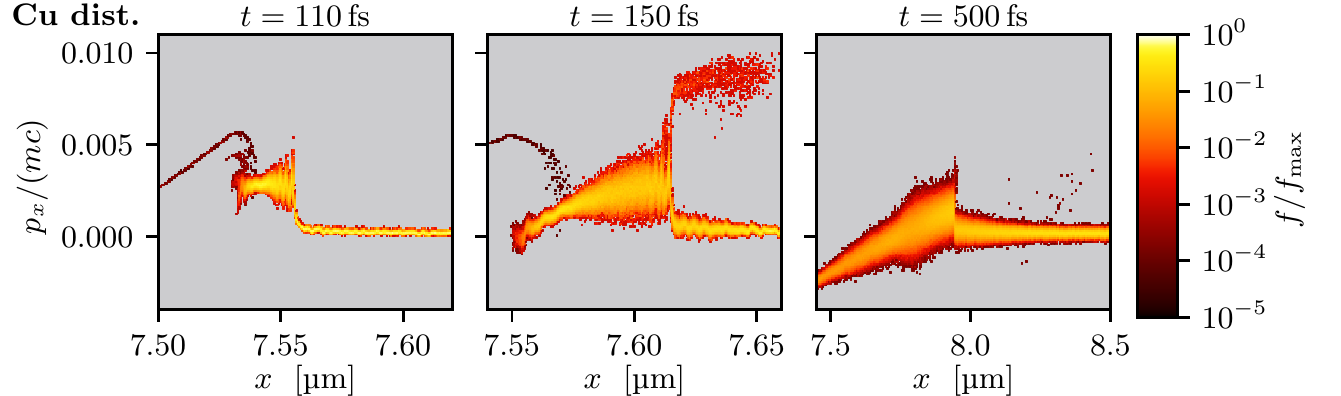}
\caption{Distribution of copper ions in a $\unit[2.5]{\micro{m}}$
  thick target, with collisions, at times $t=\unit[110]{fs}$, 
 $t=\unit[150]{fs}$ and $\unit[500]{fs}$. Note that the initial
  position of the target front is at $x=\unit[7.5]{\micro{m}}$,
  different from the other simulations presented. }
\label{fig:Cu-thick_i-dist}
\end{figure*}

\begin{figure}
\centering
\includegraphics{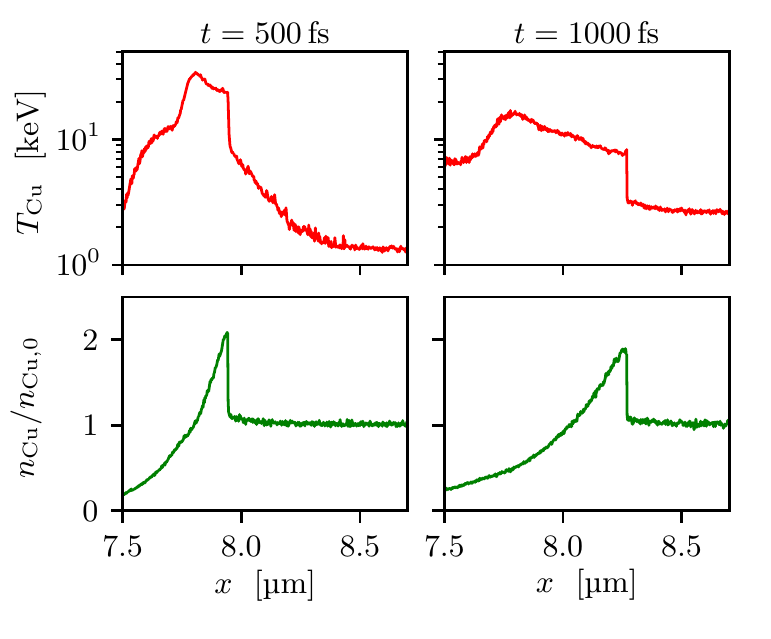}
\caption{Spatial profiles of the copper ion temperature and density at
  $t=\unit[500]{fs}$ and $t=\unit[1000]{fs}$. Both profiles display a
  sharp jump; the temperature jump by about a factor of $2.6$, while
  the density jumps by about a factor $2.0$.
}
\label{fig:Cu-thick_temp}
\end{figure}

As our final setup, we switch to a $\unit[2.5]{\micro{m}}$ copper
target, driven by an $a_{0}=10$ and $\unit[60]{fs}$ FWHM duration
pulse. Those parameters may be of interest to warm-dense-matter
studies~\cite{ElectronPaper2020}. %
The simulation results are shown in Fig.~\ref{fig:Cu-thick_i-dist}.
Despite the significant increase in target areal density, the measured
electron temperature still reaches $T_{\ee}\approx\unit[3.5]{keV}$,
thus the initial evolution of the shock is very similar to that in the
thin-target simulation. Indeed, the Mach number and shock potential
evolve as those displayed in
Fig.~\ref{fig:Cu-thin_long-laser_shock-M-dphi}, both qualitatively and
quantitatively (when accounting for a time shift of
${\simeq}\unit[15]{fs}$ corresponding to the different target
position).
The initial shock wave displays characteristic features of an
electrostatic shock, such as ion reflection and a velocity modulation
in the downstream. However, it also shows signs of a collisional
shock, such as isotropization of the downstream ion distribution
(i.e. the longitudinal and transverse temperatures are comparable to
each other).  The shock can therefore be claimed to be in a hybrid
regime between a collisionless electrostatic shock and a hydrodynamic
shock.

Since the target is now significantly thicker, the shock wave has time
to further dissipate its energy, and the ion reflection terminates at
$t\approx\unit[300]{fs}$, well before the shock front encounters the
rarefaction wave from the back of the target. %
As the shock steadily loses speed~--~and the electrostatic potential
drop across the shock front decreases~--~a point is reached when the
electrostatic potential barrier is too weak to cause ion reflection
(in fact, the electric field reaches the level of statistical
noise). However, even though the ion reflection is absent, the steady
propagation of a clear shock front structure in phase space is clearly
visible inside the target, in the $t=\unit[500]{fs}$ panel of
Fig.~\ref{fig:Cu-thick_i-dist}. There are corresponding
discontinuities in the ion temperature, $T_{\Cu}$, and density,
$n_{\Cu}$, profiles: Figure~\ref{fig:Cu-thick_temp} shows that
$T_{\Cu}$ and $n_{\Cu}$ jump by a factor of $2.6$ and $2.0$,
respectively.  Since the laser no longer exerts radiation pressure on
the target front side, the latter rapidly expands towards the vacuum
as a rarefaction wave propagates into the shocked plasma. At
$t=\unit[500]{fs}$, this rarefaction wave has caught up with the shock
front to create a weakly supersonic ($\MM\approx1.3$), planar blast
wave ~\cite[Sec.~4.3]{Drake_book2006}, which slowly decays away
(compare the ion temperature and density jumps at $t=\unit[500]{fs}$
and $t=\unit[1000]{fs}$).

To study how the qualitative features of the shock dynamics depend on
laser parameters, further simulations have been performed; $a_0$
ranging from $2$ to $14$ and the pulse FWHM duration ranging from
$\unit[15]{fs}$ to $\unit[120]{fs}$. In addition, two simulations have
been run with $a_0=7$ and $14$, with the respective pulse durations
varied such that the pulse energy would stay the same as in the case
presented above ($a_0=10$ and $\unit[60]{fs}$ FWHM duration). We could
identify two qualitatively different regimes for the ion dynamics. At
lower intensities, the pulse is not strong enough to initiate ion
reflection at any point; instead, a shock-like structure similar to
the one displayed at $t=\unit[110]{fs}$ in
Fig.~\ref{fig:Cu-thick_i-dist} is launched, and is sustained for
several hundred femtoseconds, with its speed and amplitude decaying
rather slowly. This behaviour is observed here for $a_0\le 7$, and
also in the simulation with $a_0=7$ and $\unit[120]{fs}$ FWHM
duration. The latter indicates that that both the laser intensity and
energy are important for the onset of ion reflection.

At higher intensities, the behaviour is qualitatively similar to the
one shown in Fig.~\ref{fig:Cu-thick_i-dist}~--~ion reflection is
initiated, followed by a gradual loss of energy, until ion reflection
no longer occurs, and the shock turns into a collisionally sustained
blast wave. However, the time scale for this transition to happen
depends on the laser parameters: both higher intensity and shorter
pulses result in an earlier onset of the ion reflection, as well as a
faster transition into a blast wave. The reason for the faster demise
of ion reflection may be linked to the rapid collisional heating of
the upstream ions by the reflected ions. A hotter upstream favours ion
reflection, thus hastening the shock dissipation.  Remarkably, the
ions in the downstream of the blast wave are heated to several tens of
$\unit{keV}$ temperatures, in the first ${\sim}\unit[100]{fs}$ after
the ion reflection has ended. For instance, the temperature recorded
in Fig.~\ref{fig:Cu-thick_temp} is ${\sim}\unit[20]{keV}$ at
$t=\unit[500]{fs}$. In the case of $a_{0}=14$ and FWHM duration of
$\unit[30]{fs}$, the downstream ion temperature reaches
${\sim}\unit[60]{keV}$ at $t=\unit[300]{fs}$, then dropping down to
${\sim}\unit[30]{keV}$ at $t=\unit[500]{fs}$.  Unlike the heating
scenario put forward by TSR~\cite{Turrell-etal_NCommun2015}, the
heating process of the downstream heavy ions revealed by our
simulations does not involve inter-species friction induced in the
shock electrostatic potential.

Another trend observed in the scans is that shorter duration pulses
generate faster shock evolution, i.e.\ a faster onset of ion
refection, as well as a faster decay into a blast wave. This is also
likely linked to the interaction of the laser piston and the
plasma. Shorter laser pulses are quicker to reach their maximum
intensity. The laser piston may therefore reach sufficient strength to
reflect ions, before any pre-shock perturbation (e.g.\ as that in the
$t=\unit[110]{fs}$ panel in Fig.~\ref{fig:Cu-thick_i-dist}) would have
time to form and overtake the piston. The early onset of ion
reflection then leads to a rapid transition to a blast wave, as
discussed in the previous paragraph.

In relation to the transition from hybrid shock to a blast wave, we
note that the end of the ion reflection is accompanied by an increase
in the width of the shock front, from
$\Delta x_{\sh} \sim\unit[1.6]{nm}$ (i.e., a few times the Debye
length $\lD\approx\unit[0.3]{nm}$) to $\unit[6{-}9]{nm}$. This width
is about an order of magnitude larger than the collisional ion mean
free path, here estimated as the inter-atomic distance,
$\lambda_{\rm mfp}\approx\unit[0.25]{nm}$. Our finding is consistent
with previous estimates of the width of weakly supersonic
($\MM\approx2$) hydrodynamic shocks
($\Delta{x}_{\sh}\approx20\lambda_{\rm
  mfp}$)~\cite{Bellei-etal_PoP2014,Vidal-etal_POFB1993}.

Finally, the robustness of the ion dynamics observed in the thick
copper target has been tested against possible multidimensional
effects on the laser-driven electron energization and subsequent ion
dynamics, through a two-dimensional simulation of the thick copper
target, detailed further in~\cite{ElectronPaper2020}. This simulation
reveals that the situation studied here is sufficiently collisional
that the shock does not suffer from transverse modulations.

\section{Conclusions}
Using particle-in-cell simulations, we have numerically investigated
the impact of Coulomb collisions on the ion dynamics in high-$Z^{*}$,
solid density caesium hydride and copper targets, irradiated by
high-intensity ($I\approx\unit[2{-}5\times10^{20}]{Wcm^{-2}}$),
ultrashort ($\unit[10{-}60]{fs}$), circularly polarized laser pulses.

In all cases collisional absorption through inverse Bremsstrahlung
heats the electrons up to $\unit[3{-}10]{keV}$ temperatures throughout
the target, while the use of CP reduces the creation of high-energy
electrons. Subsequently, collisions quickly relax the electrons to a
Maxwellian distribution. %
The impact of the laser pulse launches an electrostatic shock wave.
In all cases studied here, the collisionally enhanced electron heating
results in faster shock waves, with higher potential drops across the
shock front, than in the corresponding collisionless simulations.

In the CsH target, the different charge-to-mass ratios of the hydrogen
and caesium ions result in strong proton reflection. In contrast to
the results of TSR~\cite{Turrell-etal_NCommun2015}, we do not observe
a large number of protons passing through the shock front and get
heated via collisional friction with the Cs ions. Instead,
inter-species friction results in the \emph{reflected} ions being
heated up to ${\sim}\unit{keV}$ temperatures. The difference in proton
reflection between our results and those of TSR appears to be a
consequence of distinct collision models in the dense/cold plasmas
where the Spitzer theory no longer applies. This suggests that laser
plasma experiments, using targets containing a highly charged species
and protons volumetrically, may be utilized to differentiate between
numerical collision models.

In pure Cu targets, the collisional coupling between the reflected and
upstream ions is stronger, causing an appreciable heating of these
two. Also, the higher density of both ions and electrons causes a
faster decay of the shock in the CsH target.
When turning to a somewhat lower-intensity, but longer-duration laser
pulse, the initial stages of the shock launching process become more
decoupled from the laser pulse and the RPA. Here, the shock forms
already prior to the on-target laser peak. However, the shock front
continues to accelerate until about ${\sim}\unit[20]{fs}$ after the
on-target laser peak.  Because of the quick launch of the
electrostatic shock, the maximum energy of the accelerated ions has
less sharp temporal variation, since there is no transition from the
RPA ions to the CSA ions. Yet, the shock initially lacks ion
reflection, the onset of which appears to be bootstrapping itself via
heating of the upstream ions by the reflected ones.

Lastly, we increased the target thickness in order to follow the
electrostatic shock evolution over a longer duration, and to become
more relevant to high-energy-density-physics applications. While the
shock wave is at no point purely electrostatic, as it exhibits some
features of hydrodynamic shocks, we observe the shock speed and
potential drop to decay until the shock loses its capability to
reflect ions. At this stage, the electrostatic potential drop across
the shock front has also disappeared, and a rarefaction wave launched
from the target front side has overtaken the shock front, turning it
into a weakly supersonic ($\MM\approx1.3$) collisional blast
wave. This formation is capable of locally heating up the downstream
ions to tens of $\unit{keV}$ temperatures for a duration of about
${\sim}\unit[100]{fs}$.

\ack The authors are grateful for fruitful discussions with L. Hesslow
and T. F\"{u}l\"{o}p, as well as to M. Grech and F. P\'{e}rez for
support with \Smilei{}, and A. E. Turrel for providing inputs for
simulations presented in~\cite{Turrell-etal_NCommun2015}.
This project has received funding from the European Research Council
(ERC) under the European Union's Horizon 2020 research and innovation
programme under grant agreement No 647121, the Swedish Research
Council (grant no. 2016-05012), and the Knut och Alice Wallenberg
Foundation.
The simulations were performed on resources provided by the Swedish
National Infrastructure for Computing (SNIC) at Chalmers Centre for
Computational Science and Engineering (C$^3$SE) and High Performance
Computing Center North (HPC$^2$N).

\section*{\refname}
\bibliographystyle{iopart-num}
\bibliography{references}

\end{document}